\documentclass[aps,twocolumn,prl,showpacs,preprintnumbers,amsmath,amssymb]{revtex4}
\allowdisplaybreaks[4]
\newcommand{\lsim} 
 {\ \raise.35ex\hbox{$<$}\kern-0.75em\lower.5ex\hbox{$\sim$}\ }
\newcommand{\gsim}
 {\ \raise.35ex\hbox{$>$}\kern-0.75em\lower.5ex\hbox{$\sim$}\ }

\usepackage{graphicx}
\usepackage{dcolumn}
\usepackage{bm}
\usepackage{amsmath}
\usepackage{times}
\usepackage[usenames]{color}
\usepackage{natbib}

\begin{document}
\title{Theory of Valence Transition in BiNiO$_3$}
\author{Makoto Naka$^{1,2}$, Hitoshi Seo$^{1,3}$, and Yukitoshi Motome$^4$}
\affiliation{$^1$Center for Emergent Matter Science (CEMS), RIKEN, Wako 351-0198, Japan}
\affiliation{$^2$Department of Physics, Tohoku University, Sendai 980-8578, Japan}
\affiliation{$^3$Condensed Matter Theory Laboratory, RIKEN, Wako 351-0198, Japan}
\affiliation{$^4$Department of Applied Physics, University of Tokyo, 7-3-1 Hongo, Bunkyo, Tokyo, Japan}
\date{\today}
\begin{abstract} 
Motivated by the colossal negative thermal expansion recently found in BiNiO$_3$, the valence transition accompanied by the charge transfer between the Bi and Ni sites is theoretically studied. 
We introduce an effective model for Bi-$6s$ and Ni-$3d$ orbitals with taking into account the valence skipping of Bi cations, and investigate the ground-state and finite-temperature phase diagrams within the mean-field approximation. 
We find that the valence transition is caused by commensurate locking of the electron filling in each orbital associated with charge and magnetic orderings, and the critical temperature and the nature of the transitions are strongly affected by the relative energy between the Bi and Ni levels and the effective electron-electron interaction in the Bi sites. 
The obtained phase diagram well explains the temperature- and pressure-driven valence transitions in BiNiO$_3$ and the systematic variation of valence states for a series of Bi and Pb perovskite oxides. 
\end{abstract} 


\maketitle
\narrowtext



%
%

%




Perovskite transition metal (TM) oxides (general formula: $AB$O$_3$) have been providing central issues of phase transitions and strong electron correlations in condensed matter physics~\cite{Imada, Cheong}. 
They exhibit a wide range of novel magnetic, dielectric, and transport properties: for example, the large negative magnetoresistance in La$_{1-x}$Sr$_{x}$MnO$_3$~\cite{Chahara, Helmolt, Tokura}, the spin-state transition in La$_{1-x}$Sr$_{x}$CoO$_3$~\cite{Korotin, Saitoh}, the metal-to-insulator transition in $R$NiO$_3$ ($R$: rare earth element)~\cite{Torrance}, and the ferroelectric to quantum paraelectric transition in Ba$_{1-x}$Sr$_x$TiO$_3$~\cite{Sawaguchi, Zhou}. 
In these phenomena, the central players are the electrons in $3d$ orbitals of the $B$-site TMs hybridized with oxygen $2p$ orbitals.
The $A$-site cations, on the other hand, are usually inert and have been regarded as ``stagehands": 
they control the electron filling and bandwidth through their valence state and ionic radius, respectively. 

Peculiar exceptions to the above standards have recently been found in several perovskite TM oxides, in which the $A$-site cations play an active role as ``valence skipper". 
In these compounds, 
not only the $B$-site $3d$ electrons  but also the valence $s$ electrons in the $A$-site cations significantly contribute to the electronic properties. 
In the valence skippers, the outermost $s$ orbital prefers closed-shell configurations $s^{0}$ or $s^{2}$, and tends to skip the intermediate valence $s^{1}$. 
This is attributed to the effective attractive interaction between $s$ electrons~\cite{Varma, Anderson, Hase}, and hence the $A$-site valence state can be actively controlled through electronic degrees of freedom. 
Owing to the multiple electronic instabilities in both $A$- and $B$-site cations, the TM oxides with the $A$-site valence skipper have a potential of new electronic phases and functions. 

The colossal negative thermal expansion (CNTE) material BiNiO$_3$~\cite{Ishiwata_1} is one of such candidates; both Bi-$6s$ and Ni-$3d$ electrons are expected to play a key role in the large volume change~\cite{Azuma_nat, Nabetani}. 
At ambient pressure, BiNiO$_3$ has a unique valence state, where the average valence of Bi is $4+$ but it is disproportionated to $3+$ and $5+$, while the valence of Ni is $2+$~\cite{Wadati}: the $A$-site Bi cation exhibits the valence skipping nature. 
Bi${^{3+}}$ and Bi$^{5+}$ are spatially ordered in a checkerboard-like pattern, which can be regarded as a charge ordering (CO) at the Bi sites.
The electronic and lattice structures are drastically changed by applying pressure; the system exhibits an insulator-to-metal transition with the structural change from triclinic to orthorhombic, where the electrons are transferred from Ni to Bi and the valence state changes from Bi${^{3+}}_{0.5}$Bi${^{5+}}_{0.5}$Ni$^{2+}$ to Bi$^{3+}$Ni$^{3+}$. 
At the same time, the unit cell volume largely shrinks about by 5 $\%$~\cite{Azuma_nat, Ishiwata_2, Azuma_jacs}. 
Under pressure, this phase transition is also observed by raising temperature ($T$), which is termed as the CNTE. 

The large volume shrinkage in the CNTE is attributable to the valence increase at the $B$ sites which determine the lattice parameters in perovskites in general. 
However, the mechanism of the valence transition behind it, involving both Bi and Ni sites and the charge transfer between them, remains to be clarified. 
This is owing to the fact that, as mentioned above, most of the previous studies for perovskite TM oxides have been generally focusing on the TM $3d$  and oxygen $2p$ electrons, which usually govern the electronic properties, and hardly incorporate the valence state in the $A$-site cations explicitly. 

In this Letter, we present a microscopic theory for the valence transition behind the CNTE in BiNiO$_3$, which treats the active electronic degrees of freedom in both $A$- and $B$-site cations on equal footing. 
We introduce a simple but realistic effective model for BiNiO$_3$, and clarify the ground-state and finite-$T$ phase diagrams within the mean-field (MF) approximation. 
We will show that the valence transition is controlled by the relative energy between Bi and Ni levels as well as the electron correlation at the Bi sites. 
The charge transfer between Bi and Ni is caused by spontaneous symmetry breaking associated with a bipolaronic CO at Bi and antiferromagnetic (AFM) order at Ni, both of which favor a commensurate filling in each band. 
Our result will provide a useful guide for further exploration of larger CNTE in related materials. 

\begin{figure}[t]
\begin{center}
\includegraphics[width=1.0\columnwidth, clip]{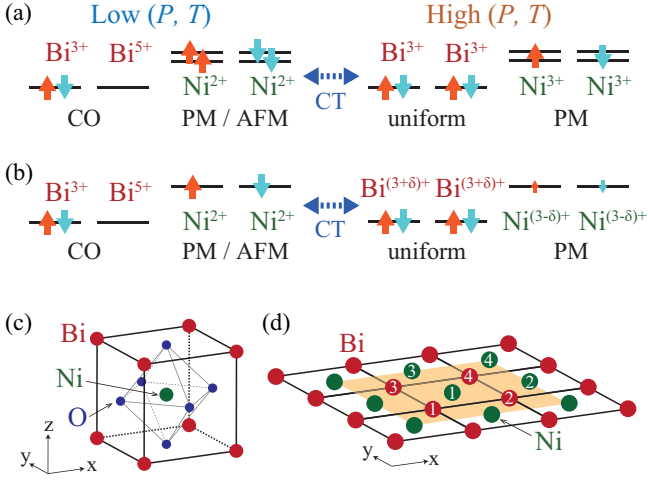}
\end{center}
\caption{(Color online) 
(a) Schematic electron configurations for Bi $6s$ and Ni $3d$ $e_{g}$ orbitals in BiNiO$_3$ and (b) those in the effective model in Eq.~\ref{eq:hamil}. 
CO, PM, AFM, and CT denote charge order, paramagnetic, antiferromagnetic states, and charge transfer, respectively. 
Schematic pictures of (c) pervskite structure for BiNiO$_3$ and (d) the projected two-dimensional lattice structure for the effective model. 
The shaded area represents the mean-field unit cell. 
}
\label{fig:vst}
\end{figure}
%
%
%
Let us construct an effective model for BiNiO$_3$. 
According to first principles band calculations, the Bi-$6s$ and Ni-$3d$ $e_{g}$ bands are located near the Fermi energy, and the O-$2p$ bands strongly hybridize with them~\cite{Azuma_jacs}. 
Here, we take into account the Bi-$6s$ and Ni-$3d$ $e_{g}$ orbitals, while assuming that the role of O-$2p$ is effectively incorporated in the energy levels and the other parameters. 
The electronic configurations in the insulating Bi${^{3+}}_{0.5}$Bi${^{5+}}_{0.5}$Ni$^{2+}$O${^{2-}}_{3}$ and the metallic Bi$^{3+}$Ni$^{3+}$O${^{2-}}_{3}$ phases are schematically illustrated in Fig.~\ref{fig:vst}(a). 
For further simplicity, we omit the twofold degeneracy of $e_{g}$ orbitals as shown in Fig.~\ref{fig:vst}(b); the essential physics related to the valence transition will be retained as described below. 
Considering the uniform electron configuration along the $z$ axis in the real compound~\cite{Ishiwata_1}, we adopt a two-dimensional lattice of Bi and Ni obtained by projecting the perovskite structure in Fig.~\ref{fig:vst}(c) on the $xy$ plane, as shown in Fig.~\ref{fig:vst}(d). 
The Hamiltonian is given by 
\begin{align}
{\cal H} &= t_{\rm N} \sum_{\langle ij \rangle \sigma}^{\rm Ni-Ni} \left( a^{\dagger}_{i \sigma} a_{j \sigma} + {\rm H.c.} \right) + t_{\rm B} \sum_{\langle ij \rangle \sigma}^{\rm Bi-Bi} \left( b^{\dagger}_{i \sigma} b_{j \sigma}+ {\rm H.c.} \right) \notag \\
&+ t_{\rm BN} \sum_{\langle ij \rangle \sigma}^{\rm Bi-Ni} \left( a^{\dagger}_{i \sigma} b_{j \sigma} + {\rm H.c.} \right) \notag \\ 
&+ {\Delta} \sum_{i \sigma}^{\rm Ni} n^{\rm N}_{i \sigma} + U_{\rm N} \sum_{i \sigma}^{\rm Ni} n^{\rm N}_{i \uparrow} n^{\rm N}_{i \downarrow} + U_{\rm B} \sum_{i \sigma}^{\rm Bi} n^{\rm B}_{i \uparrow} n^{\rm B}_{i \downarrow} \notag \\
&+ V_{\rm B} \sum_{\langle ij \rangle}^{\rm Bi-Bi} n^{\rm B}_{i} n^{\rm B}_{j} + V_{\rm BN} \sum_{\langle ij \rangle}^{\rm Bi-Ni} n^{\rm N}_{i} n^{\rm B}_{j}, 
\label{eq:hamil}
\end{align}
where $a_{i \sigma}$ and $b_{i \sigma}$ represent the annihilation operators of electron with the spin $\sigma(=\uparrow, \downarrow)$ at the Ni and Bi sites of $i$-th unit cell, respectively; 
$n^{\rm N}_{i \sigma} = a^{\dagger}_{i \sigma} a_{i \sigma}$ and $n^{\rm B}_{i \sigma} = b^{\dagger}_{i \sigma} b_{i \sigma}$. 
In Eq.~(\ref{eq:hamil}), the first and second lines represent the electron hopping on the Ni-Ni, Bi-Bi, and Bi-Ni bonds. 
In the third line, the first term is the energy difference between the Bi-$6s$ and Ni-$3d$ levels, the second term is the on-site Coulomb interactions on the Ni sites, and the third term is the effective interaction that describes the valence skipping nature of Bi~\cite{Varma, Anderson, Hase}; we consider not only positive but also negative values for $U_{\rm B}$. 
The fourth line represents the intersite Coulomb interactions on the Bi-Bi and Bi-Ni bonds. 

We obtain the phase diagram for the model in Eq.~(\ref{eq:hamil}) by the MF approximation with decoupling the two-body interaction terms in the third and fourth lines in Eq.~(\ref{eq:hamil}) as $n n' \simeq n \langle n' \rangle + \langle n \rangle n' - \langle n \rangle \langle n' \rangle$. 
We take the unit cell that includes four Bi and four Ni sites, as shown in Fig.~\ref{fig:vst}(d). 
By investigating several sets of parameters and varying them, we find that $U_{\rm B}$ and $\Delta$ are most relevant to the valence transition; hence, below we show the results by varying these two parameters while fixing the others at $t_{\rm N} = t_{\rm B} = 1$, $t_{\rm BN} = 0.5$, $U_{\rm N} = 3$, $V_{\rm B} = 0.65$, $V_{\rm BN} = 1$. 
We have confirmed that the detailed changes of the parameters do not alter the following results qualitatively. 

\begin{figure}[t]
\begin{center}
\includegraphics[width=1.0\columnwidth, clip]{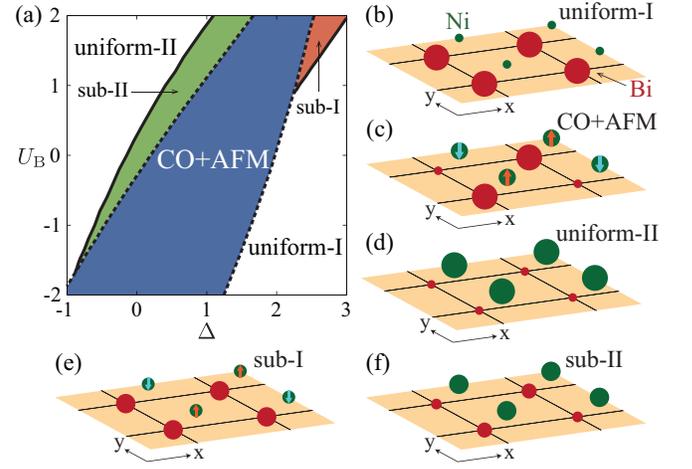}
\end{center}
\caption{(Color online) 
(a) Ground state phase diagram. 
Solid and broken lines denote the second and first order transitions, respectively. 
Schematic pictures of charge and spin configurations: (b) uniform-I, (c) CO+AFM, (d) uniform-II, (e) sub-I, and (f) sub-II. 
The circles and arrows represent the charge densities and spin moments at each site, respectively. 
}
\label{fig:latt_pdgs}
\end{figure}
%
%
%
\begin{figure}[t]
\begin{center}
\includegraphics[width=1.0\columnwidth, clip]{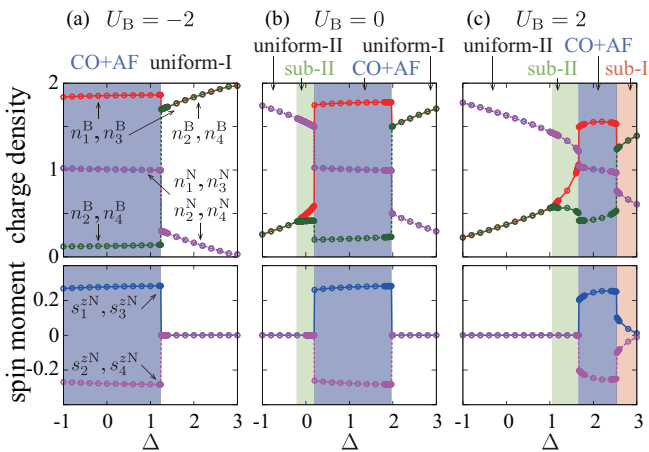}
\end{center}
\caption{(Color online) 
$\Delta$ dependences of the charge densities (upper panels) and spin moments (lower panels) at each Bi and Ni site for $t_{\rm B} = t_{\rm N} =1$, $t_{\rm BN} = 0.5$, $U_{\rm N} = 3$, and $V_{\rm B} = 0.65$; (a) $U_{\rm B} = -2$, (b) $U_{\rm B} = 0$, and (c) $U_{\rm B} = 2$. 
}
\label{fig:op_gs}
\end{figure}
%
%
%
First, we present the ground state properties. 
Figure~\ref{fig:latt_pdgs}(a) shows the obtained phase diagram on the plane of $\Delta$ and $U_{\rm B}$. 
There are three main phases with different valence states [Figs.~\ref{fig:latt_pdgs}(b)--\ref{fig:latt_pdgs}(d)], while two sub-phases appear between them [Figs.~\ref{fig:latt_pdgs}(e) and \ref{fig:latt_pdgs}(f)]. 
In the uniform-I (II) phase, a major portion of the electrons reside in the Bi (Ni) sites with spatially uniform distribution, and no spin polarization is seen in either sites as shown in Fig.~\ref{fig:latt_pdgs}(b) [\ref{fig:latt_pdgs}(d)]. 
On the other hand, in the CO+AFM phase, the average charge density becomes almost the same for Bi and Ni, i.e., charge transfer occurs between them compared to the uniform phases. 
In addition, a bipolaronic charge disproportionation occurs in a staggered way between the Bi sites, as shown in Fig.~\ref{fig:latt_pdgs}(c). 
The spin moments are canceled at each Bi site, whereas those at the Ni sites are antiferromagnetically ordered. 

To elucidate variations of the electronic states in more detail, we show the $\Delta$ dependences of the charge densities $\langle n^{\rm B(N)}_{i} \rangle = \langle n^{\rm B(N)}_{i \uparrow} \rangle +  \langle n^{\rm B(N)}_{i \downarrow} \rangle$ and spin moments $\langle s_{i}^{z {\rm N}} \rangle = (\langle n^{\rm N}_{i \uparrow} \rangle - \langle n^{\rm N}_{i \downarrow} \rangle)/2$. 
The spin moments in the Bi sites remain zero in the parameter range we calculated. 
Figure~\ref{fig:op_gs}(a) shows the results in $U_{\rm B} = -2$, where the valence skipping nature of Bi is relatively strong. 
In the large-$\Delta$ region, where the energy level of the Ni orbital is substantially higher than that of Bi, 
the electrons occupy dominantly and uniformly the Bi sites, and the Ni orbitals are nearly empty; no spin polarization appears. 
This electronic state corresponds to the uniform-I phase in Fig.~\ref{fig:latt_pdgs}(b). 
When $\Delta$ is decreased, the charge densities and the Ni spin moments simultaneously jump at $\Delta \simeq 1.25$, with showing the spontaneous symmetry breaking in both charge and spin channels: the system enters the CO+AFM phase. 
At the transition, the average charge density in the Bi (Ni) sites suddenly decreases (increases) by about $0.7$. 
Namely, the valence transition is caused by the charge transfer between Bi and Ni. 

As seen in the phase diagram in Fig.~\ref{fig:latt_pdgs}(a), when $U_{\rm B}$ is increased, the CO+AFM phase shrinks and shifts to larger $\Delta$ region. 
This is seen in Figs.~\ref{fig:op_gs}(b) and \ref{fig:op_gs}(c), showing the results for $U_{\rm B} = 0$ and $U_{\rm B} = 2$, respectively. 
When $U_{\rm B} = 0$, the uniform-II phase is realized in the small-$\Delta$ region, where the charge density in the Ni sites becomes larger than that in the Bi sites, on the contrary to the uniform-I phase. 
The transition between the CO+AFM and uniform-II phases is also a valence transition. 
In addition, a narrow intermediate state termed sub-II [Fig.~\ref{fig:latt_pdgs}(f)] appears, in which the average charge densities are similar to the uniform-II phase but with a weak checkerboard-type CO at the Bi sites. 
For $U_{\rm B} = 2$, 
the region of $\Delta$ for the CO+AFM phase is further narrowed and shifted, 
and another intermediate phase termed sub-I [Fig.~\ref{fig:latt_pdgs}(e)] appears between the CO+AFM and uniform-I phase, where a weak AFM order occurs at the Ni sites. 
We note that a larger $U_{\rm B}$ results in a smaller amplitude of the valence changes, as shown in Figs.~\ref{fig:op_gs}(a)--\ref{fig:op_gs}(c). 

\begin{figure}[t]
\begin{center}
\includegraphics[width=0.8\columnwidth, clip]{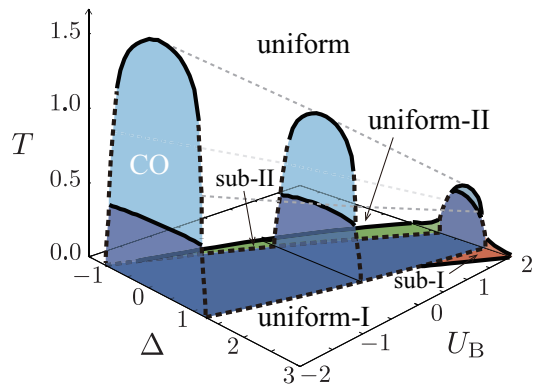}
\end{center}
\caption{(Color online) 
Global phase diagram in the $\Delta$-$U_{\rm B}$-$T$ space. 
Solid and broken lines denote the second and first order transitions, respectively. 
Thin dotted lines are guides for the eye denoting connections between the CO domes. 
}
\label{fig:pd_global}
\end{figure}
%
%
%
Next, we discuss the finite-$T$ properties. 
Figure~\ref{fig:pd_global} shows the global phase diagram with the $T$-axis attached to the ground-state phase diagram in Fig.~\ref{fig:latt_pdgs}(a). 
When $T$ is raised in the CO+AFM phase,  first the AFM order is destroyed, while the CO remains; at higher $T$ the CO melts and the uniform phase is realized. 
The uniform-I and -II phases in the ground state are continuously connected in the high $T$ region, as they possess the same symmetry. 
Consequently, the CO phase shows a dome-like structure. 
The width of the CO dome becomes wider for lower $T$, whose rate is slightly increased below the AFM transition temperature. 

\begin{figure}[t]
\begin{center}
\includegraphics[width=1.0\columnwidth, clip]{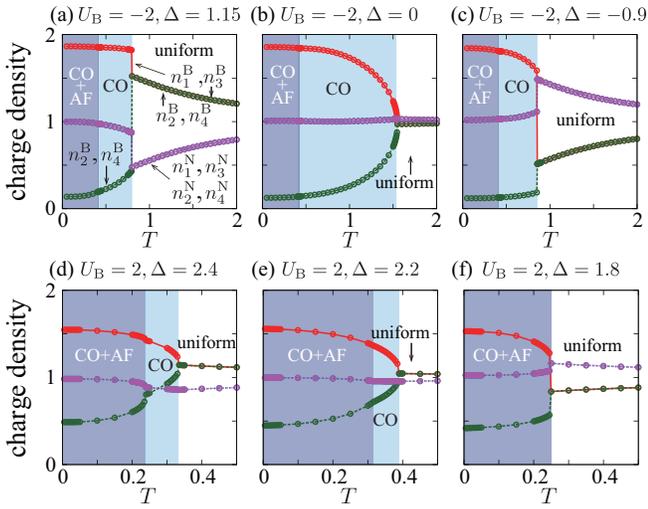}
\end{center}
\caption{(Color online) 
$T$ dependences of the charge densities at each Bi and Ni site for $t_{\rm B} = t_{\rm N} =1$, $t_{\rm BN} = 0.5$, $U_{\rm N} = 3$, and $V_{\rm B} = 0.65$; (a)--(c) $U_{\rm B} = -2$ and (d)--(f) $U_{\rm B} = 2$ for different values of $\Delta$ (indicated in the figures). 
Note the difference in the $T$ range for the two choices of $U_{\rm B}$. 
}
\label{fig:op_ft}
\end{figure}

The $T$ dependences of the charge densities at each site are shown in Figs.~\ref{fig:op_ft}(a)--\ref{fig:op_ft}(c) for several values of $\Delta$ at $U_{\rm B} = - 2$. 
The CO transition is discontinuous for the large and small values of $\Delta$, while it becomes continuous for intermediate $\Delta$. 
The discontinuous CO transition is accompanied by the large amount of the charge transfer between the Bi and Ni sites. 
Below the CO transition temperature, the average charge densities in the Bi and Ni sites become nearly $1$. 
The transition between CO and CO+AFM is always continuous, whose critical temperature is insensitive to $\Delta$. 
Figures~\ref{fig:op_ft}(d)--\ref{fig:op_ft}(f) show the results for $U_{\rm B} = 2$; 
all the $T$ scales become smaller compared to the case of $U_{\rm B} = -2$, and accordingly, the amplitude of the CO also becomes smaller. 
The average charge densities in the Bi and Ni sites change even below the AFM transition temperature, which moderately depends on $\Delta$ here, in contrast to the case of $U_{\rm B} = -2$. 

%
%
%
Through the above analyses, we have found that the valence transition occurs between the uniform and CO phases. 
Here, we discuss its origin. 
The Bi-site electrons have the instability toward the bipolaronic CO due to the combination of negative $U_{\rm B}$ and positive $V_{\rm B}$. 
The instability is most enhanced when the Bi band is half-filled; $\sum_{i} \langle n_{i}^{\rm B} \rangle/ N = 1$ (the sum is taken over the Bi sites and $N$ is number of the Bi sites). 
This condition is satisfied near the center of the CO dome where the CO temperature is maximized, for example in the situation in Fig.~\ref{fig:op_ft}(b). 
On the other hand, apart from the dome center, the Bi band filling becomes incommensurate in the high-$T$ uniform phase, as shown in Figs.~\ref{fig:op_ft}(a) and \ref{fig:op_ft}(c). 
In this case, the CO transition is accompanied by the charge transfer between the Bi and Ni sites for restoring the commensurability of the Bi band. 
This commensurate locking is the primary cause of the valence transition. 
The AFM instability in the Ni sites due to the positive $U_{\rm N}$ also contributes to this locking effect, as clearly seen in Fig.~\ref{fig:op_ft}(d). 
These considerations lead us to conclude that the valence transition is attributed to the commensurate locking of the electron filling in both Bi and Ni bands driven by the strong electron correlations. 

Finally, let us compare the present results with the experiments in BiNiO$_3$ and related materials. 
The first-order valence transition with the charge transfer from Ni to Bi is observed by increasing $T$ and pressure in BiNiO$_3$. 
In the present calculation, the discontinuous valence transition driven by $T$ is indeed seen, e.g., in Fig.~\ref{fig:op_ft}(a). 
The pressure-driven transition can also be explained by the transition with increasing the relative orbital energy $\Delta$, e.g., in Fig.~\ref{fig:op_gs}(a), when we assume that the pressure affects $\Delta$. 
This is reasonable as the volume reduction under pressure is mainly due to the contraction of the Ni-O perovskite framework in BiNiO$_3$~\cite{Azuma_nat}, which may increase $\Delta$ via the increase of the Ni $3d$ energy levels.

On the other hand, $\Delta$ can be more directly controlled by the substitution of the TM cations. 
Such effects are actually investigated in Bi$M$O$_3$ ($M=$ V, Cr, Mn, Fe, Co, and Ni)~\cite{BiCr, BiMn, BiFe, BiCo} and Pb$M$O$_3$ ($M=$ Ti, V, Cr, Fe, and Ni)~\cite{PbCr, PbNi}, where Pb is another valence skipper favoring the oxidation states Pb$^{2+}$($6s^{2}$) and Pb$^{4+}$($6s^{0}$). 
In the Bi compounds, the valence state is Bi$^{3+}M^{3+}$ for $M=$ V -- Co, while it is Bi${^{3+}}_{0.5}$Bi${^{5+}}_{0.5}M^{2+}$ for $M=$ Ni (at ambient pressure) as we have been discussing. 
Meanwhile, in Pb$M$O$_3$, successive variations of such valence state as Pb$^{2+}M^{4+}$ $\leftrightarrow$ Pb${^{2+}}_{0.5}$Pb${^{4+}}_{0.5}M^{3+}$ $\leftrightarrow$ Pb$^{4+}M^{2+}$ are observed 
between V and Cr, and Fe and Ni, respectively, 
where the intermediate state Pb${^{2+}}_{0.5}$Pb${^{4+}}_{0.5}M^{3+}$ involves the CO of Pb$^{2+}$ and Pb$^{4+}$ resembling Bi${^{3+}}_{0.5}$Bi${^{5+}}_{0.5}M^{2+}$. 
These experimental results suggest that the electrons tend to be accumulated in the $A$($B$) cations in the systems with low (high) $3d$ level, while they are uniformly distributed in $A$ and $B$ cations for the intermediate case. 
This tendency and the phase transitions between the different valence states are well understood by the $\Delta$ dependence in our phase diagram in Fig.~\ref{fig:pd_global}. 

In summary, we have presented and analyzed a microscopic model for the electronic states in the perovskite oxides including the valence skipper as the $A$-site cation, especially focusing on the valence transition in BiNiO$_3$. 
We have found that the valence transition is attributed to the commensurate locking of the electron filling in the Bi and Ni bands due to the electron correlations, and it is sensitive to the relative energy of the Bi $6s$ and Ni $3d$ levels and the intra Bi-site interaction. 
Our work provides a fundamental understanding of the electronic properties in a series of new perovskite materials including valence skippers as the $A$-site cation which have as yet been scarcely dealt with. 
As a mechanism of negative thermal expansion, the ion radius change due to the valence transition is markedly distinct from those previously discussed, which are attributed to lattice vibrations and magnetovolume effects~\cite{Takenaka}. 
We expect that our result will contribute to the development of the ``next generation" negative thermal expansion materials. 

The authors would like to thank M. Azuma, S. Ishihara, T. Mizokawa, M. Mizumaki, K. Oka, and T. Watanuki for valuable discussions. 


\end{document}